\documentclass{article}

\usepackage[T1]{fontenc}
\usepackage{graphicx}%
\usepackage{multirow}%
\usepackage{amsmath,amssymb,amsfonts}%
\usepackage{mathrsfs}%
\usepackage[title]{appendix}%
\usepackage{xcolor}%
\usepackage{textcomp}%
\usepackage{manyfoot}%
\usepackage{booktabs}%
\usepackage{algorithm}%
\usepackage{algorithmicx}%
\usepackage{algpseudocode}%
\usepackage{listings}%
\usepackage{textcomp}%
\usepackage{url}

\title{Multimodal Learning for Scalable Representation of High-Dimensional \\Medical Data}

\author{Areej Alsaafin$^1$,  Abubakr Shafique$^1$,  Saghir Alfasly$^1$, \\ Krishna R. Kalari$^2$, H.R.Tizhoosh$^1$ \\
\vspace{0.1in}\\
Kimia Lab, Dept. of Artificial Intelligence \& Informatics,\\ Mayo Clinic, Rochester, MN, USA \and Division of Computational Biology, \\Dept. of Quantitative Health Sciences,\\  Mayo Clinic, Rochester, MN, USA}

\begin{document}
\maketitle
\begin{abstract}
Integrating artificial intelligence (AI) with healthcare data is rapidly transforming medical diagnostics and driving progress toward precision medicine. However, effectively leveraging multimodal data, particularly digital pathology whole slide images (WSIs) and genomic sequencing, remains a significant challenge due to the intrinsic heterogeneity of these modalities and the need for scalable and interpretable frameworks. Existing diagnostic models typically operate on unimodal data, overlooking critical cross-modal interactions that can yield richer clinical insights. We introduce MarbliX (Multimodal Association and Retrieval with Binary Latent Indexed matriX), a self-supervised framework that learns to embed WSIs and immunogenomic profiles into compact, scalable binary codes, termed ``monogram.'' By optimizing a triplet contrastive objective across modalities, MarbliX captures high-resolution patient similarity in a unified latent space, enabling efficient retrieval of clinically relevant cases and facilitating case-based reasoning.
\textcolor{black}{In lung cancer, MarbliX achieves 85–89\% across all evaluation metrics, outperforming histopathology (69–71\%) and immunogenomics (73–76\%). In kidney cancer, real-valued monograms yield the strongest performance (F1: 80–83\%, Accuracy: 87–90\%), with binary monograms slightly lower (F1: 78–82\%).}

\end{abstract} 
  
\section{Introduction}\label{introduction}
Cancer diagnosis has traditionally relied on expert pathologists manually examining tissue slides under a microscope. While molecular testing has improved diagnostic precision in recent years, morphological assessment remains a manual and labor-intensive task. The rise of artificial intelligence (AI), particularly large-scale models, is beginning to shift this paradigm by uncovering complex, high-dimensional patterns in clinical data, enabling more integrative and data-driven cancer diagnostics. In parallel, progress in cancer immunogenomics has underscored the critical role of the adaptive immune system in identifying and eliminating tumor cells. T and B lymphocytes respond to tumor-specific antigens, and similarities in T cell receptor (TCR) and B cell receptor (BCR) sequences reveal shared antigenic responses across patients~\cite{gun2019targeting,beausang2017t,pogorelyy2018exploring}. These patterns facilitate patient stratification and therapeutic targeting~\cite{sidhom2021deeptcr,medzhitov1997innate}, with immune repertoire diversity metrics linked to differential treatment outcomes~\cite{pogorelyy2018exploring,jung2004unraveling}. Such insights help explain heterogeneous responses among clinically similar patients~\cite{fridman2012immune,leone2020metabolism} and support improved prediction of treatment efficacy and resource allocation~\cite{alsaafin2023deep}.

Recent deep learning models have effectively analyzed immunogenomic data for tasks such as outcome prediction, receptor clustering, and neoantigen discovery~\cite{sidhom2021deeptcr,beshnova2020novo,burger2018targeting,fischer2020predicting}. Concurrently, histopathological imaging provides a rich morphological view of tumors, offering a complementary perspective to molecular data. Each modality captures distinct biological signals, and their integration holds promise for a more holistic understanding of disease. However, the heterogeneity and high dimensionality of these data types present significant challenges for joint modeling. Manual interpretation is infeasible at scale, and existing computational tools typically address only a single modality, limiting their utility. Patient heterogeneity further highlights the need for computational models that support personalized tumor characterization~\cite{alizadeh2015toward}. Multimodal learning offers a compelling approach, yet frameworks that effectively unify imaging and immunogenomic data remain scarce. \textcolor{black}{The goal of this work is to determine whether a compact, binary multimodal representation—jointly learned from WSIs and immunogenomic data—can reliably preserve clinically relevant patient similarity for retrieval and subtype characterization.}
\begin{figure}
    \vspace{-23pt}
    \setlength{\abovecaptionskip}{-15pt} 
    \setlength{\belowcaptionskip}{-10pt}  
    \centering
    \includegraphics[width=1\textwidth]{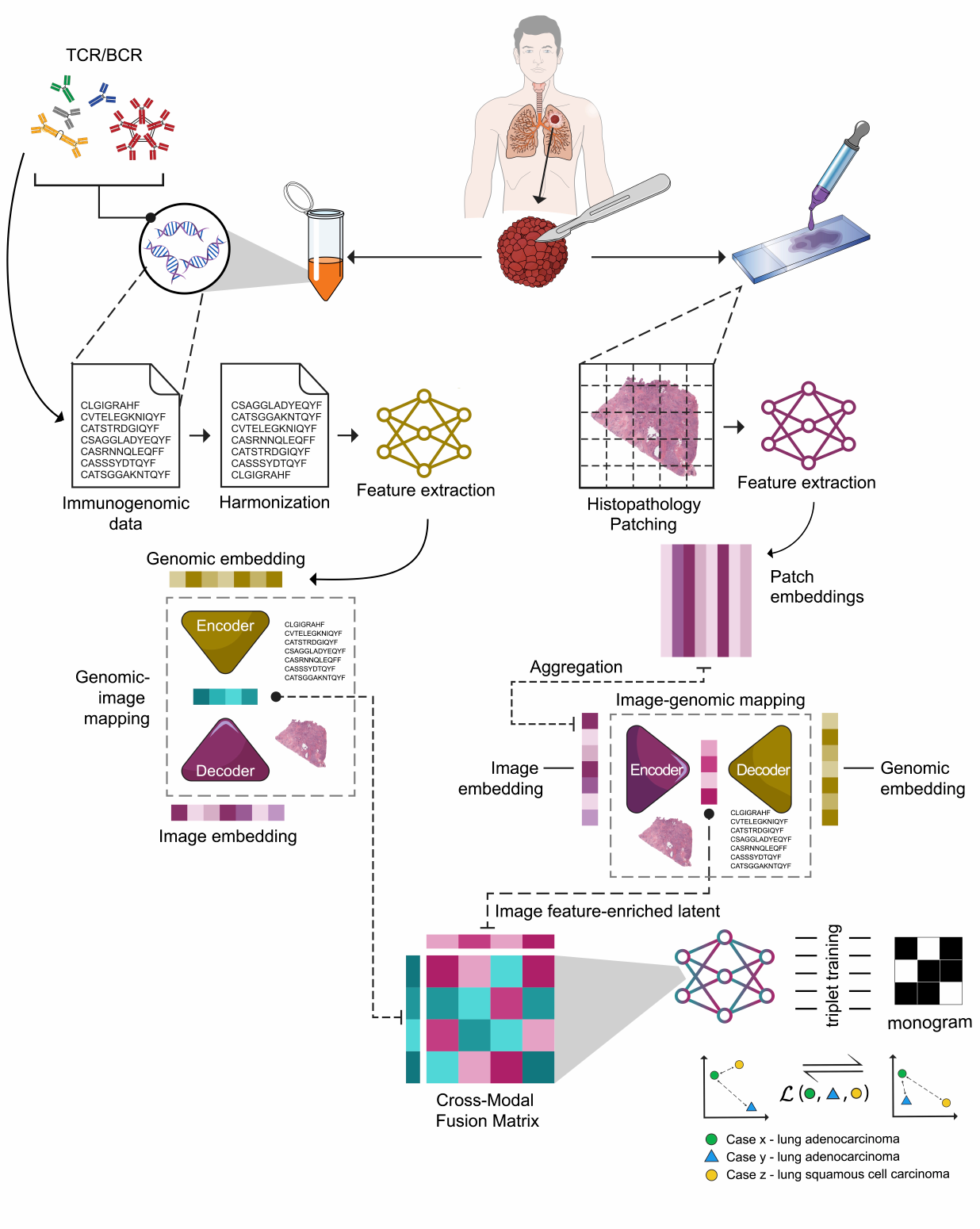}
    \caption{MarbliX integrates histopathology images and immunogenomic data to generate personalized binary multimodal representations using self-supervised triplet contrastive learning.}
    \label{fig:marblix}
\end{figure}

\textcolor{black}{
\subsection{Related Works}
\textbf{Multimodal learning with histopathology and omics} -- A growing body of work integrates histopathology with molecular data for prognosis and representation learning \cite{alsaafin2023learning,chen2021histopathological,vincenzo2024review,waqas2024digital}. Pathomic Fusion and related frameworks \cite{chen2020pathomic} combine WSI-derived features with genomics and clinical variables using late fusion or attention-based modules for diagnosis and survival prediction, demonstrating that complementary omics signals can improve performance over image-only models. More recently, TANGLE  \cite{jaume2024transcriptomics} proposes transcriptomics-guided slide representation learning: modality-specific encoders for WSIs and gene expression are aligned via a contrastive objective to produce joint slide embeddings that are effective for few-shot classification and retrieval. 
These methods, however, typically operate on continuous gene expression vectors, assume well-aligned expression–slide pairs, and focus on downstream classification or survival tasks rather than compact, indexable patient codes. Moreover, they do not address immune repertoire–level sequence data, which differ substantially in structure and sparsity from bulk transcriptomics.}

\textcolor{black}{\textbf{Cross-modal transformers and multimodal pretraining in pathology} -- Transformer-based multimodal models have been introduced to jointly reason over WSIs and non-image information \cite{raza2025ps3}. MCAT \cite{chen2021multimodal} uses a co-attention transformer to fuse slide-level representations with clinical/genomic covariates for survival prediction, treating the multimodal problem as weakly supervised MIL at gigapixel scale. 
 GECKO \cite{kapse2025gecko} pretrains a dual-branch MIL network by aligning WSI embeddings with an interpretable ``concept prior'' derived from textual pathology descriptors, and can optionally incorporate transcriptomics when available. 
 These approaches highlight the power of multimodal pretraining and attention-based fusion but are architecturally heavy, assume dense patch sets or concept maps, and ultimately produce real-valued high-dimensional slide embeddings. None of them are designed to yield ultra-compact binary representations for large-scale indexing, nor do they target WSI–immunogenomic repertoire integration.
Beyond pathology, large-scale vision–language models such as CLIP, e.g., CONCH \cite{lu2024visual}, align images and text via contrastive training on hundreds of millions of pairs, enabling zero-shot transfer and cross-modal retrieval in natural-image domains. 
 While conceptually related at the level of learning a shared latent space, these models rely on rich natural language supervision and internet-scale paired data, conditions that do not hold for WSIs and immune-repertoire sequences.}
 
\textcolor{black}{\textbf{Cross-modality retrieval} -- Cross-modal retrieval methods generally focus on mapping heterogeneous modalities—most commonly image–text —into a shared embedding space where similarity can be measured directly. LILE \cite{maleki2024self} is a dual-attention transformer network for cross-modal retrieval in histopathology archives, aligning image and text modalities into a shared latent space. 
 It augments standard cross-attention with an additional self-attention loss term that enriches intra-modal representation before cross-modal matching. 
Proceedings of Machine Learning Research
 On benchmark datasets such as MS-COCO and ARCH, LILE outperforms prior cross-modal methods, demonstrating more accurate information retrieval between images and text. 
 Its design highlights the value of attention-based alignment for cross-modality tasks, but — like most such methods — it produces high-dimensional continuous embeddings rather than compact binary codes, which limits scalability for large-scale archives.}
 
\textcolor{black}{\textbf{Compact and binary representation learning} -- 
Compact and binary codes have a long history in large-scale image retrieval, where hashing methods map visual features into Hamming space to enable efficient storage and approximate nearest-neighbour search. 
 In medical imaging and pathology, hashing-based approaches have been proposed to generate low-dimensional binary representations of histopathology images for fast retrieval in large archives. Yottixel \cite{kalra2020yottixel} introduced the hashing of patch-level embeddings into compact ``barcodes,'' \cite{tizhooshminmax} and the aggregation of these into a ``bunch of barcodes'' \cite{tizhoosh2024image} for each WSI, enabling lean indexing and fast retrieval using Hamming distance. Other methods \cite{hemati2023learning} frame barcode generation as a combinatorial optimization problem and use an evolutionary algorithm to find a permutation of features that yields more discriminative barcodes. Applied across medical and non-medical datasets (including pathology images), this method significantly improves retrieval accuracy compared to arbitrary feature orderings. 
 While these methods demonstrate the practicality of binary codes for efficient search, they are almost exclusively unimodal (image-only) and do not address the challenge of jointly encoding heterogeneous biomedical signals (e.g., morphology and immunogenomics) into a single compact structure.}
 
\textcolor{black}{\textbf{Limitations of existing work for WSI–sequence integration and scalability} -- 
Taken together, existing multimodal frameworks in computational pathology establish that combining WSIs with molecular or clinical data can improve prediction and sometimes retrieval. 
 However, they typically: 1) Focus on continuous transcriptomic features rather than immune-repertoire–level sequence data, 2) Produce high-dimensional real-valued embeddings rather than binary codes designed for indexing and storage efficiency, and 3) Rely on architectures (co-attention transformers, dense MIL pretraining) whose computational and memory footprints make them less suitable as core indexing mechanisms in very large archives.}
 
\textcolor{black}{In contrast, the present work targets a different point in this design space: learning compact binary monograms that jointly encode WSI morphology and immunogenomic information, with the explicit goal of enabling scalable, retrieval-oriented patient representations rather than optimizing a single supervised prediction task.}

\subsection{Contributions}
In this paper, we introduce MarbliX (Multimodal Association and Retrieval with Binary Latent Indexed matriX), a novel multimodal framework that bridges histopathology and immune receptor sequencing. MarbliX employs self-supervised representation learning to embed whole slide images (WSIs) and immune repertoires into a shared binary latent space. These compact embeddings, termed ``monogram'', encode both morphological and immunogenomic patterns, enabling efficient similarity retrieval across patients to support case-based reasoning. As shown in Figure~\ref{fig:marblix}, MarbliX allows clinicians to query a patient and retrieve similar cases based on integrated multimodal evidence. Unlike existing search tools that operate solely on WSIs or their subregions~\cite{kalra2020yottixel,Lahr2024,tizhoosh2024image}, MarbliX provides a richer and more personalized representation. \textcolor{black}{This approach strengthens diagnostic relevance and interpretability within a research setting, while laying the groundwork for scalable multimodal decision-support tools that could ultimately translate into clinical practice.}

\textcolor{black}{MarbliX makes four key contributions: First, it introduces the monogram, a compact representation that encapsulates diverse patient data into a single binary signature. These monograms summarize patterns across modalities, facilitating downstream tasks such as personalized diagnostics and case comparison. Second, MarbliX compresses large datasets—including WSIs and sequencing data—into binary barcodes. This compactness improves efficiency in storage, computation, and processing, making the framework scalable for large-scale applications such as search and retrieval. Third, its backbone-independent design allows MarbliX to integrate heterogeneous data sources by mapping them into a shared latent space. This flexibility enhances model generalization and adaptability to various biomedical data types. Finally, MarbliX supports monogram-based search, enabling retrieval of similar cases through direct comparison of patient monograms. This functionality enhances diagnostic support and clinical research by identifying patients with shared characteristics.}
\section{Methods}
This section describes the details of MarbliX's design and the intricate process involved in generating a unique \emph{monogram} representation. An overview of MarbliX is illustrated in Figure~\ref{fig:marblix}. 
The design of MarbliX involves three main phases: unimodal transformation, multimodal latent association, and monogram representation. The details of every phase is described below. 

\subsubsection*{Unimodal Transformation}
The integration of histopathology images and immune cell sequencing data into a shared computational framework requires an alignment step to transform them into a common format. This enables joint manipulation and integration within a unified model. Hence, as a first stage, each modality, is processed and transformed into a single feature vector or embedding. For simplicity, in the context of the notation $(\textbf{\textit{I}}, S)$, $\textbf{\textit{I}}$ represents the histopathology image, while $S$ represents the immunogenomic data (a set of immune cell sequences) of a given case.\\
\noindent\textbf{Image processing}: to represent the WSIs, SPLICE~\cite{alsaafin2024splice} was employed to select representative patches, forming a \emph{collage} for image $\textbf{\textit{I}}$ after segmenting the tissue region from the background using Otsu thresholding. This collage is a condensed representation, composed of a select set of representative patches extracted from $\textbf{\textit{I}}$, capturing the crucial tissue characteristics that define the image. The \emph{collage} was generated by setting the similarity threshold to the 30th percentile, striking a balance between performance and computational/storage requirements. Once the \emph{collage} is generated for $\textbf{\textit{I}}$, the next step involves extracting deep features from the individual patches that compose the \emph{collage}. This process is achieved by leveraging a pre-trained deep neural network $\mathcal{F}$, here we used DINO ViT~\cite{caron2021emerging}, which possesses the ability to extract patterns and meaningful information within these patches. We chose DINO ViT for its strong self-supervised visual representation capabilities, which generalize well even in domains such as histopathology despite not being domain-specific. As a result of this indexing, a feature vector $\textbf{\textit{f}}_i$ of patch $\textbf{\textit{P}}_i$ within the collage is generated by applying $\textbf{\textit{f}}_i = \mathcal{F}(\textbf{\textit{P}}_i)$, where $\textbf{\textit{f}}_i \in \mathbb{R}^{l \times 1}$. 
To craft an all-encompassing feature vector that encapsulates the entirety of image $\textbf{\textit{I}}$ and effectively represents its rich content, a widely adopted practice involves computing the average of the patch-level feature vectors~\cite{vale2021long}, resulting in a single feature vector $\textbf{\textit{f}} \in \mathbb{R}^{l \times 1}$ that serves as a holistic representation of the entire image $\textbf{\textit{I}}$. As we used DINO ViT, this resulted in a 768-dimensional embedding for the entire WSI.

\noindent\textbf{Sequencing data processing}: for the immunogenomic data, raw RNA-seq files from TCGA were utilized to reconstruct the immune repertoire of every patient. TRUST4~\cite{song2021trust4} was employed to obtain the TCR and BCR sequences of each patient from their RNA-seq profiles. 
\textcolor{black}{Rare sequences were filtered globally across the entire dataset based on overall frequency thresholds, not within subtype classes}
Through experimentation, for the lung dataset, sequences that were not common to at least 30\% of the patients within the subtype class were excluded, while a lower threshold of 15\% was applied to kidney cases due to the limited number of samples. Before we encode the sequencing data into a dense vector, we applied Seqwash~\cite{alsaafin2024harmonizing} method to preprocess the sequencing profiles and prepare them for feature extraction. Seqwash is a ``harmonization'' approach tailored to genetic sequencing data and serves as a crucial preprocessing step, aimed at preparing these sequences for analysis using deep models designed for textual data by overcoming the impact of the variability among patients in terms of sequence lengths and unregulated sequence orders. Therefore, Seqwash was used to unify the patient profiles by aligning them into a standardized representation before proceeding with deep feature extraction. The ultimate goal is to create a single, coherent embedding that encapsulates vital information while negating the effects of varying sequence orders within each patient's profile. Applying Seqwash on sequencing profile $S$ results in a harmonized set $S_h$. A pre-trained deep learning model $\mathcal{G}$ is then employed to distill features from the sequences by applying $\textbf{\textit{g}} = \mathcal{G}(S_h)$, where $\textbf{\textit{g}} \in \mathbb{R}^{l \times 1}$ represents a feature vector extracted from the harmonized sequences set $S_h$. Here, we employed BERT which resulted in a 768-dimensional embedding.

\subsubsection*{Multimodal Latent Association}
Following the transformation of each modality, histopathology image $\textbf{\textit{I}}$ into embedding $\textbf{\textit{f}}$ and immune cell sequence profile $S$ into embedding $\textbf{\textit{g}}$, the association between these embeddings is learned. However, as these embeddings come from different models, they have different ranges. Therefore, min-max rescaling was performed to bring the embeddings to a common scale before learning the association between them. After normalization, association learning was performed by projecting the two embeddings into a shared latent space. In this shared space, the embeddings from both modalities coalesce to form a concise patient representation. This step addresses the issue of non-relevant features that may exist in the uni-modal data obtained from a pretrained network. By merging these embeddings into an encoded representation, we want to extract and consolidate the pertinent features from each modality, enhancing their combined synergy and informative value in subsequent analyses.\\
To accomplish this, as shown in Algorithm~\ref{algo:blended_latent}, two deep neural networks with an encoder-decoder (autoencoder) architecture are employed, each tailored to emphasize the salient features from its corresponding modality while suppressing extraneous information. This step addresses the issue of non-relevant features that may exist in the uni-modal data obtained from a pretrained network. By merging these embeddings into an encoded representation, we want to extract and consolidate the pertinent features from each modality, enhancing their combined synergy and informative value in subsequent analyses. This is achieved by training the two hybrid models to generate an encoded latent representation, highlighting the relevant features. Specifically, each autoencoder model is designed to take one modality and reconstruct the other, resulting in a latent representation that embodies the dominant features of its respective modality. \textcolor{black}{Reconstructing one modality from another using hybrid autoencoders tests whether the two data types share meaningful, learnable structure. If tissue and omics can predict each other, the model uncovers latent biological signals that transcend any single modality. This cross-reconstruction forces the network to learn aligned representations rather than modality-specific noise. It also provides a built-in check on multimodal coherence, revealing when information is missing, redundant, or biologically disconnected.} 

\begin{algorithm}
\caption{Multimodal Latent Association}
\label{algo:blended_latent}
\begin{algorithmic}[1]
\State \textbf{Input:}
\State \quad Histopathology image embedding $\textbf{\textit{f}}$
\State \quad Immune cell sequence profile embedding $\textbf{\textit{g}}$
\State \textbf{Training Stage:} \label{line:image_encoder_start}
\State \quad Initialize $\mathcal{A}_I$ (Autoencoder for Histopathology):
\State \quad \textbf{while} not converged \textbf{do}:
\State \quad \quad Forward pass: $\textbf{\textit{f}} \rightarrow \mathcal{E}_I(\textbf{\textit{f}}) \rightarrow \mathcal{D}_I(\mathcal{E}_I(\textbf{\textit{f}}))$
\State \quad \quad Compute loss: $l_I = \text{MSE}(\textbf{\textit{g}}, \mathcal{D}_I(\mathcal{E}_I(\textbf{\textit{f}})))$
\State \quad \quad Backpropagate and update weights \label{line:image_encoder_end}
\State \quad Initialize $\mathcal{A}_S$ (Autoencoder for Immune Cell Sequences): \label{line:seq_encoder_start}
\State \quad \textbf{while} not converged \textbf{do}:
\State \quad \quad Forward pass: $\textbf{\textit{g}} \rightarrow \mathcal{E}_S(\textbf{\textit{g}}) \rightarrow \mathcal{D}_S(\mathcal{E}_S(\textbf{\textit{g}}))$
\State \quad \quad Compute loss: $l_S = \text{MSE}(\textbf{\textit{f}}, \mathcal{D}_S(\mathcal{E}_S(\textbf{\textit{g}})))$
\State \quad \quad Backpropagate and update weights \label{line:seq_encoder_end}
\State \textbf{Latent Representation:} \label{line:blend_start}
\State \quad Encode using the encoder of $\mathcal{A}_I$:
\State \quad \quad $\textbf{\textit{u}} \leftarrow \mathcal{E}_I(\textbf{\textit{f}})$
\State \quad Encode using the encoder of $\mathcal{A}_S$:
\State \quad \quad $\textbf{\textit{v}} \leftarrow \mathcal{E}_S(\textbf{\textit{g}})$
\State \textbf{return} $\textbf{\textit{u}}$, $\textbf{\textit{v}}$ \label{line:blend_end}
\end{algorithmic}
\end{algorithm}

In the first stage (illustrated in Algorithm~\ref{algo:blended_latent}, Lines \ref{line:image_encoder_start}-\ref{line:image_encoder_end}), an autoencoder denoted as $\mathcal{A}_I$ is designed, where it takes the histopathology image embedding $\textbf{\textit{f}}$ as input, and reconstructs the immune cell sequence profile embedding $\textbf{\textit{g}}$. During training, $\mathcal{A}_I$ learns to focus on critical features present in histopathology images that offer insights into immune cell sequence patterns. These features are encapsulated within the bottleneck layer, situated just before the first decoder layer of model $\mathcal{A}_I$. Conversely, in the second stage (described in Algorithm~\ref{algo:blended_latent}, Lines \ref{line:seq_encoder_start}-\ref{line:seq_encoder_end}), another autoencoder, denoted as $\mathcal{A}_S$, is employed, which takes the immune cell sequence profile embedding $\textbf{\textit{g}}$ as input and reconstructs the histopathology image embedding $\textbf{\textit{f}}$. This design empowers the model to emphasize essential characteristics inherent to immune cell sequence data that are relevant to the histopathological context, also embedded within the bottleneck layer of model $\mathcal{A}_S$.

Both hybrid autoencoder models $\mathcal{A}_I$ and $\mathcal{A}_S$ comprised an encoder with two dense layers of size 512 and 256, a bottleneck layer of size 128, and a decoder with two dense layers of size 256 and 512. All models were trained using Adam optimizer and mean square error (MSE) as the loss function. The image-genomics autoencoder was trained for 150 epochs with a learning rate of $1\times10^{-5}$, while the genomics-image autoencoder was trained for 50 epochs with a learning rate of $1\times10^{-4}$. Following the training of $\mathcal{A}_I$ and $\mathcal{A}_S$, the encoder from each hybrid autoencoder is employed to generate an encoded latent for each sample, resulting in two encoded vectors: image features-enriched latent and genomic features-enriched latent (Algorithm~\ref{algo:blended_latent}, Lines \ref{line:blend_start}-\ref{line:blend_end}). For instance, $\textbf{\textit{u}}$, characterized by a strong emphasis on histopathological features, is derived through the application of $\textbf{\textit{u}} = \mathcal{E}_I(\textbf{\textit{f}})$, where $\textbf{\textit{u}} \in \mathbb{R}^{l \times 1}$. Likewise, $\textbf{\textit{v}}$, accentuating immune cell characteristics, is derived as $\textbf{\textit{v}} = \mathcal{E}_S(\textbf{\textit{g}})$, where $\textbf{\textit{v}} \in \mathbb{R}^{l \times 1}$. The resulting compact representation (size 128) not only reduces dimensionality but also encapsulates critical aspects of both modalities. This representation serves as a powerful encoding of joint information extracted from histopathology images and immune cell sequences, facilitating profound integration in the subsequent phase.
\subsubsection*{MarbliX Monogram}
After generating the two latent representations $\textbf{\textit{u}}$ and $\textbf{\textit{v}}$, the next crucial step within the MarbliX framework is the projection and indexing of these representations into a 2D binary matrix, referred to as ``monogram.'' This matrix serves as a compact  representation to capture the intricate relationships and correlations that exist between the modalities. The process of learning the representation of \emph{monogram} is the core of the MarbliX framework, allowing a comprehensive exploration of the joint information encoded in $\textbf{\textit{u}}$ and $\textbf{\textit{v}}$. To accomplish this task, a deep neural network, denoted as $\mathcal{Q}$, is designed to uncover the correlations between histopathology and immunogenomic features based on diagnosis. This process aims to unveil the underlying structure that interlinks histopathological characteristics with immune cell behavior among cases within the same diagnostic class. In other words, the model is engineered to capture the commonality in multimodal relationships among cases that share similar diagnoses, embedding these features within their respective \emph{monograms}. Simultaneously, it strives to discern the distinctions between cases of different diagnostic classes and accentuate these disparities within their corresponding monograms. This is achieved by employing self-supervised training using triplet loss to minimize the distance between patients with the same primary diagnosis and maximize the distance between patients with different primary diagnoses. \\
As described in Algorithm~\ref{algo:patient_matrix_learning}, The $\mathcal{Q}$ model comprises three branches with shared weights, each taking a pair of latent representations, \(\textbf{\textit{u}}\) and \(\textbf{\textit{v}}\). Thus, the model is provided with triplet pairs as input 
$\mathcal{Q}(\{\textbf{\textit{u}}, \textbf{\textit{v}}\}, \{\textbf{\textit{u}}^+, \textbf{\textit{v}}^+\}, \{\textbf{\textit{u}}^-, \textbf{\textit{v}}^-\}), $
consisting of an anchor pair $(\textbf{\textit{u}}, \textbf{\textit{v}})$, a positive pair $(\textbf{\textit{u}}^+, \textbf{\textit{v}}^+)$, and a negative pair $(\textbf{\textit{u}}^-, \textbf{\textit{v}}^-)$. The anchor case serves as the reference for which the model endeavors to generate a representative monogram. The positive case shares the same diagnosis as the anchor, reinforcing common features. In contrast, the negative case differs in diagnosis from the anchor, shedding light on the discrepancies between diagnostic classes. The positive and negative samples were selected for each anchor sample by calculating pairwise Euclidean distances and identifying the farthest positive sample and the closest negative sample for each anchor. This approach was implemented to ensure robust training by introducing hard triplets to the model, guiding toward learning the similarities between samples belonging to the same class, despite eventual dissimilarity between them. Analogously, this approach guides the model to generate different representations for cases that share common features in their data but belong to different classes.
\begin{algorithm}
\caption{Learning Multimodal Monogram Representation}
\label{algo:patient_matrix_learning}
\begin{algorithmic}[1]
\State \textbf{Input:}
\State \quad $\textbf{\textit{u}}$: image latent representation  
\State \quad $\textbf{\textit{v}}$: sequencing latent representation 
\State \textbf{Training Stage:}
\State \quad Initialize $\mathcal{Q}$ model with three branches $\mathcal{T}_1$, $\mathcal{T}_2$, $\mathcal{T}_3$
\State \quad $\mathcal{T}_1$, $\mathcal{T}_2$, $\mathcal{T}_3 \leftarrow \text{shared weights}$
\State \quad \textbf{for each triplet} $\mathcal{T}_1(\textbf{\textit{u}}, \textbf{\textit{v}})$, $\mathcal{T}_2(\textbf{\textit{u}}^+, \textbf{\textit{v}}^+)$, $\mathcal{T}_3(\textbf{\textit{u}}^-, \textbf{\textit{v}}^-)$ \textbf{do}:
\State \quad \quad $M \leftarrow \textbf{\textit{u}} \otimes \textbf{\textit{v}}$ 
\State \quad \quad $\bar{M} \leftarrow \mathcal{T}(M) \leftarrow M$
\State \quad \quad $\bar{M}_{\text{binary}} \leftarrow \{ \textbf{if } w > 0.5 \textbf{ then } 1 \textbf{ else } 0 \textbf{ for each } w \textbf{ in } \bar{M} \}$
\State \quad Calculate triplet loss: \label{line:triplet_start}
\State \quad \quad $d(\textbf{\textit{a}}, \textbf{\textit{p}}) \leftarrow \text{distance}(\bar{M}, \bar{M}^+)$
\State \quad \quad $d(\textbf{\textit{a}}, \textbf{\textit{n}}) \leftarrow \text{distance}(\bar{M}, \bar{M}^-)$
\State \quad \quad $\mathcal{L}_{\text{triplet}} \leftarrow \max \{(d(\textbf{\textit{a}}, \textbf{\textit{p}}) - d(\textbf{\textit{a}}, \textbf{\textit{n}}))+ \alpha, 0\}$
\State \quad $\mathcal{T}_1$, $\mathcal{T}_2$, $\mathcal{T}_3 \leftarrow$ update weights using gradient descent \label{line:triplet_end}
\end{algorithmic}
\end{algorithm}
Within each branch (Algorithm~\ref{algo:patient_matrix_learning}, Lines \ref{line:triplet_start}-\ref{line:triplet_end}), the pair of \(\textbf{\textit{u}}\) and \(\textbf{\textit{v}}\) is projected into a matrix through the computation of the outer product between their respective layers. This tensor is then flattened and passed through three consecutive dense layers of size 1024, 256, and 64 to learn the deep multimodal relationship. The last layer of the model has a binary branch that generates a binary representation of the last layer. This is crucial as it enables the generation of compact binary representations, highly efficient for subsequent indexing and storage. As the tanh function results in values within the range of $[-1, 1]$, the binary dense layer sets positive values to 1 and negative values to 0. \\
The triplet loss function  
\begin{equation}
\label{eq:triplet}
    \mathcal{L}_{\text{triplet}}(\textbf{\textit{a}}, \textbf{\textit{p}}, \textbf{\textit{n}}) = \max \{d(\textbf{\textit{a}}, \textbf{\textit{p}}) - d(\textbf{\textit{a}}, \textbf{\textit{n}}) + \alpha, 0\}
\end{equation}
calculates the distances between the anchor's predicted matrix and both the positive and negative matrices. Here, \(\textbf{\textit{a}}\) represents the anchor case, \(\textbf{\textit{p}}\) signifies a positive case (sharing the same diagnosis as the anchor), and \(\textbf{\textit{n}}\) denotes a negative case (with a different diagnosis from the anchor). \(d(\textbf{\textit{a}}, \textbf{\textit{p}})\) calculates the distance between the anchor and positive case matrices, while \(d(\textbf{\textit{a}}, \textbf{\textit{n}})\) computes the distance between the anchor and negative case matrices. The margin parameter \(\alpha\) ensures a minimum separation between the positive and negative cases.\\
Model $\mathcal{Q}$ was trained for 150 epochs using the \emph{tanh} activation function and Adam optimizer with a learning rate of $1\times10^{-5}$. After training, the $\mathcal{Q}$ model,  generates binary monogram representations for new cases. This was done by applying $\text{monogram} = \mathcal{Q}(\{\textbf{\textit{u}}, \textbf{\textit{v}}\})$, where monogram represents an $8\times8$ binary matrix (with encoding capability to cover $2^{64}=1.8\times10^{19}$ combinations) derived from the latent representations $\textbf{\textit{u}}$ and $\textbf{\textit{v}}$ of the  image and immune cell sequences, respectively. \textcolor{black}{The monogram—implemented as a small binary matrix—is intentionally designed for lean storage when indexing hyperdimensional bimodal data. It serves as an internal representation only; the user, such as a pathologist, does not need to view or interpret it.}

\textcolor{black}{MarbliX’s novelty lies in converting heterogeneous multimodal patient data (histopathology + immunogenomics) into a compact binary “monogram” code via a unified latent association framework, enabling efficient patient matching and retrieval across modalities.}

\section{Results}
\begin{figure}
    \vspace{-15pt}
    \setlength{\abovecaptionskip}{0pt} 
    \setlength{\belowcaptionskip}{-5pt} 
    \centering
\includegraphics[width=\textwidth]{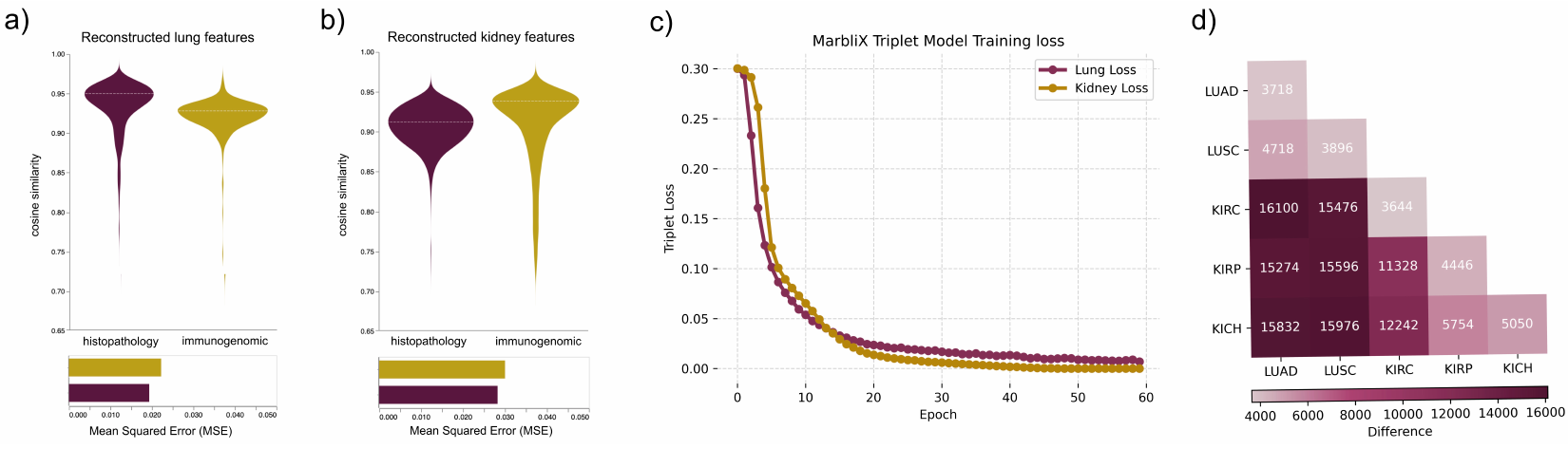}
\caption{(a–b) Violin plots of cosine similarity between original and reconstructed embeddings from hybrid autoencoders. (c) Triplet loss during MarbliX training to learn multimodal patient codes ``monograms''. (d) Heatmap showing XOR-based dissimilarities between monograms.}\label{fig:results1}
\end{figure}
MarbliX was implemented and evaluated using histopathology and genomic data from The Cancer Genome Atlas (TCGA), focusing on two primary sites: \textbf{lung} and \textbf{kidney}. Only cases with both WSIs and genomic profiles were included. The lung dataset comprised 535 lung adenocarcinoma (LUAD) and 510 lung squamous cell carcinoma (LUSC) cases. The kidney dataset included 508 kidney renal clear cell carcinoma (KIRC), 248 kidney renal papillary cell carcinoma (KIRP), and 38 kidney chromophobe (KICH) cases. Evaluation was performed using 5-fold cross-validation for the lung data and 2-fold cross-validation for the kidney data due to the limited KICH samples, ensuring adequate representation from all subtypes for training. \textcolor{black}{All datasets were divided into stratified folds at the patient level to ensure that no slide or immunogenomic data from the same patient appeared in both training and testing. For triplet construction, we used a standard approach: within each training fold, anchor–positive pairs were formed from patients sharing the same diagnosis label, while negatives were drawn from patients with different labels; all sampling occurred only within the training fold to avoid leakage. Triplets were refreshed each epoch to increase sampling diversity.}

The implementation and experiments were conducted on a Linux-based server with two AMD EPYC 7413 CPUs and four NVIDIA A100 GPUs (80GB each). The GPUs were used exclusively for unimodal data processing (feature extraction). For histopathology images, this step took approximately 7–12 hours per dataset, depending on sample size, while feature extraction from immune repertoire sequences was significantly faster (~4-9 minutes per dataset). All MarbliX training and evaluation were performed on the CPU, as the model operates on latent representations and does not require GPU acceleration. Training with 5-fold cross-validation took about 3 minutes per fold (~15 minutes per dataset). The reported experiments accounted for the majority of the computational cost, with exploratory analyses and hyperparameter tuning contributing an additional ~8 hours of overhead. \textcolor{black}{That feature extraction constitutes the main computational burden of MarbliX; however, this preprocessing is a one-time expense. Once the unimodal encoders are trained, MarbliX achieves scalability through compact binary monograms that enable rapid indexing, storage, and retrieval across very large archives. After this initial step, processing new WSIs is fast because SPLICE selects only a small set of representative patches. Regarding training on CPUs, our framework is agnostic to compute hardware, and the CPU-based training was chosen to demonstrate practicality rather than impose a limitation.}

\textcolor{black}{We should bear in mind that TCGA WSIs suffer from some limitations, including variable staining, scanner differences, tissue folds, pen marks, frozen-section artefacts, and inconsistent annotation quality. Many slides also contain degraded or poorly sectioned tissue and heterogeneous preprocessing pipelines that amplify noise. Models trained on such data frequently may learn spurious visual cues instead of true pathology. This may lead to inflated benchmark performance, poor robustness to distribution shift, and weak generalization to real clinical workflows. Despite these limitations, TCGA remains the largest publicly accessible multimodal pathology dataset, making it indispensable for baseline benchmarking and methodological development.}
\subsection*{MarbliX Training Evaluation}
Several experiments assessed the quality of patient representations. One evaluated the multimodal latent associations learned via hybrid autoencoders that map between histopathological and immunogenomic features. Figure \ref{fig:results1} shows cosine similarity between original and reconstructed embeddings—histopathology and immunogenomic—using pretrained and trained autoencoders. Violin plots for test cases in (a) lung and (b) kidney datasets show high median similarities (0.95 for lung histopathology, 0.91 for kidney; 0.93 for lung immunogenomics, 0.94 for kidney), indicating strong feature retention. Tightly packed quartiles reflect consistent performance, though a wider spread in kidney immunogenomics suggests reconstruction challenges due to limited KICH data. The bar plot in Figure \ref{fig:results1} quantifies reconstruction quality via mean squared error (MSE), which stays consistent (0.020–0.030) across both modalities and datasets. Training curves in Figure \ref{fig:results1}(c) show loss decreasing over epochs with triplet loss; both lung and kidney models converge, though the kidney model shows slower reduction, possibly due to greater data variability. To assess intra- and inter-subtype variation, binary monograms were compared using bitwise XOR on 19 test cases per subtype. The heatmap in Figure \ref{fig:results1}(d) shows lower intra-subtype dissimilarity for LUAD, LUSC, and KIRC than for KIRP and KICH, likely due to limited training data. KIRP and KICH appear more similar to each other than to KIRC, suggesting shared features or underrepresentation. Lung monograms are more similar to each other than to kidney monograms, and kidney subtypes show greater inter-subtype dissimilarity.

\begin{figure}
    \centering
    \vspace{-15pt}
    \setlength{\abovecaptionskip}{5pt} 
    \setlength{\belowcaptionskip}{-10pt} 
\includegraphics[width=0.9\textwidth,height=\textheight,keepaspectratio]{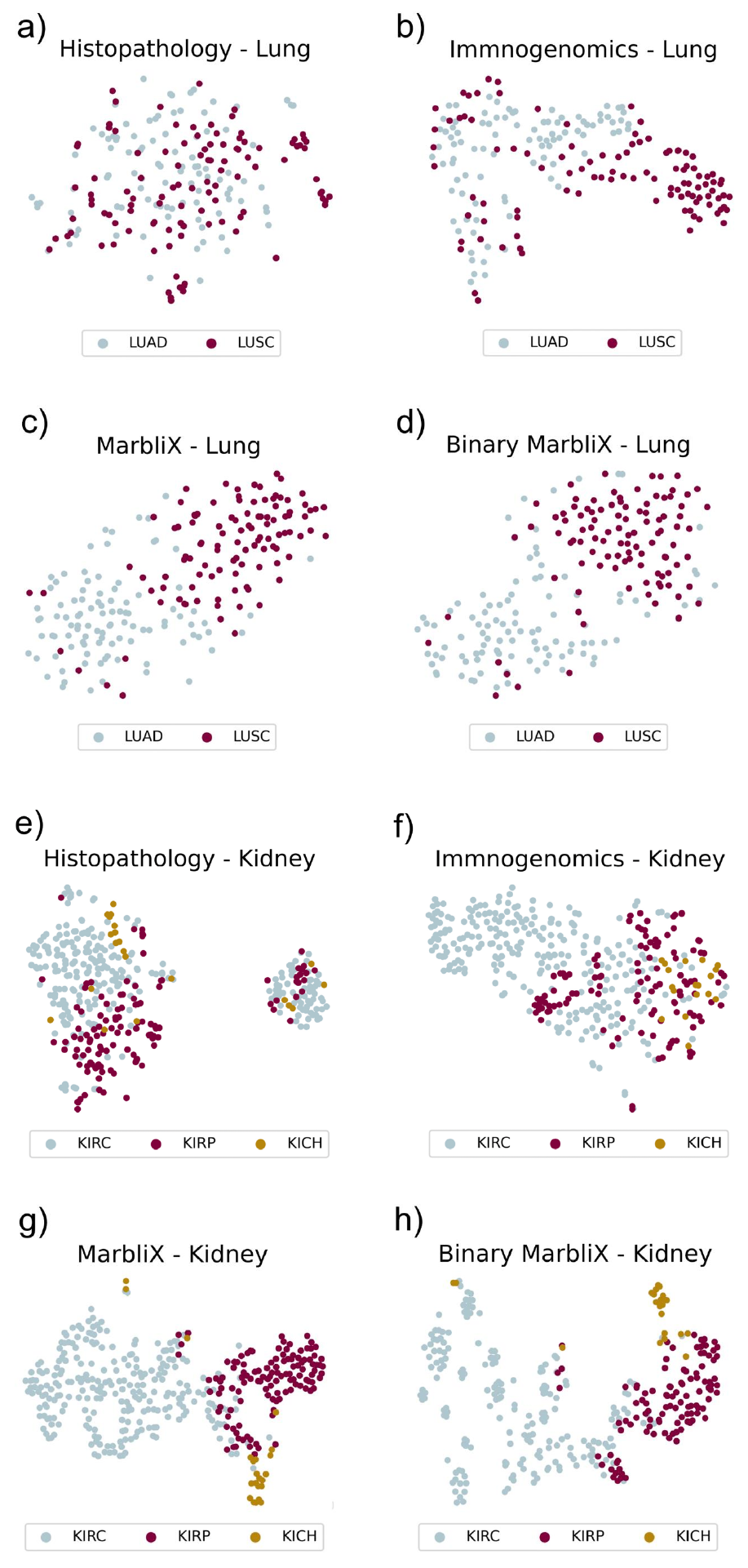}
        \caption{t-SNE maps illustrating the distribution of modalities in a reduced high-dimensional space, following PCA to 64 components. (a, e) Image embeddings; (b, f) immunogenomics; (c, g) real-valued monograms learned by MarbliX; (d, h) binary monograms learned by MarbliX.}
        \label{fig:tsne}
\end{figure}

\subsection*{MarbliX Generates Discriminative Patient Representations}
To evaluate MarbliX’s ability to generate effective multimodal representations, we compared unimodal embeddings from histopathology and immunogenomics data. PCA extracted the top 64 components from test sets (unseen folds) of lung and kidney data, followed by t-SNE projection (Figure \ref{fig:tsne}). In Figure \ref{fig:tsne}(a), LUAD and LUSC histopathology embeddings show substantial overlap, while immunogenomics (Figure \ref{fig:tsne}(b)) offers partial separation. MarbliX improves class separation, especially with binary monograms (Figures \ref{fig:tsne}(c) and (d)). For kidney data, Figure \ref{fig:tsne}(e) shows clusters influenced by hospital-specific imaging artifacts. Immunogenomics (Figure \ref{fig:tsne}(f)) better separates subtypes, with KIRC most distinct. MarbliX further enhances subtype separability (Figures \ref{fig:tsne}(g), (h)), demonstrating its ability to produce compact, discriminative representations.

\begin{figure}
    \vspace{-30pt}
    \setlength{\abovecaptionskip}{0pt} 
    \centering        \includegraphics[width=\textwidth,height=\textheight,keepaspectratio]{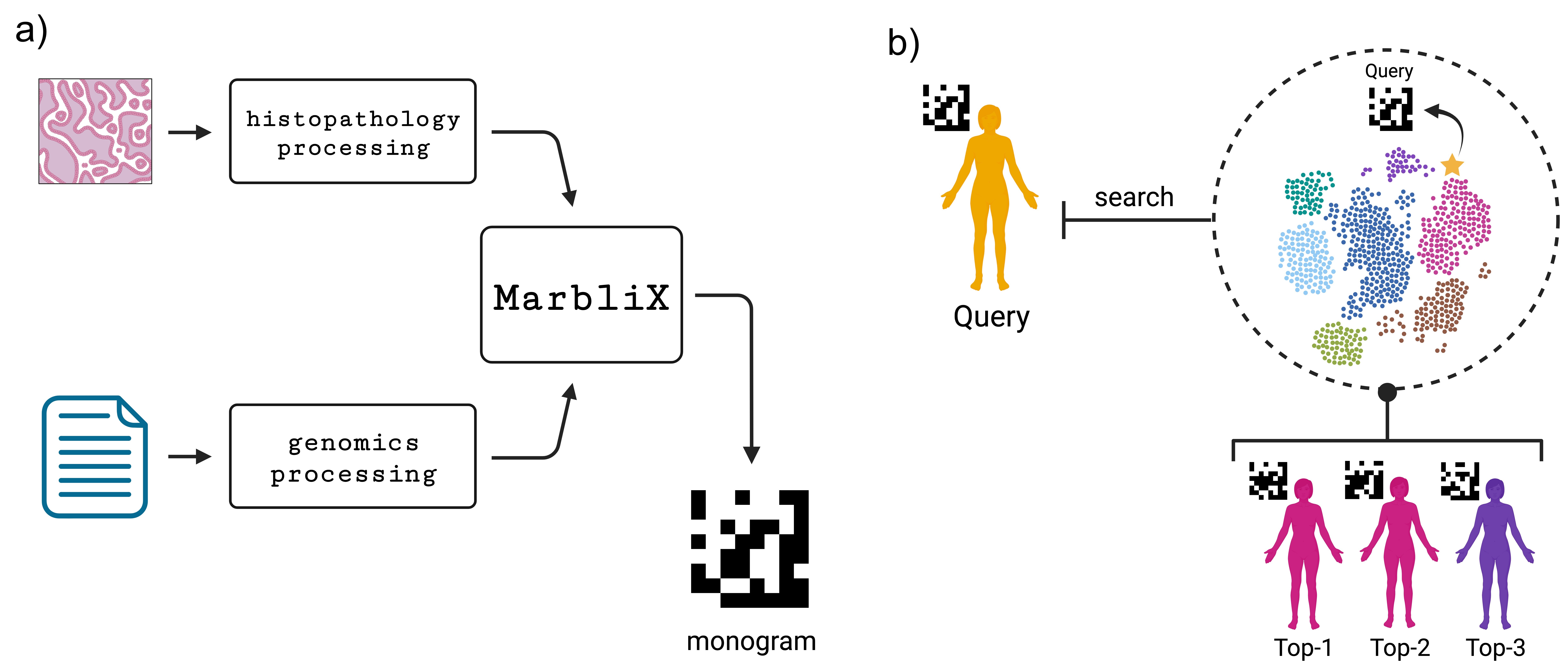}
       \caption{(a) MarbliX transforms each modality and fuses them into a monogram representation, (b) which is used to search a biomedical archive via Hamming distance to retrieve similar cases.}
       \label{fig:multimodal_search}
\end{figure}

\subsection*{Efficient Multimodal Similarity Search}
\begin{figure}
    \centering
    \setlength{\abovecaptionskip}{-5pt} 
    \setlength{\belowcaptionskip}{-25pt} 
        \includegraphics[width=\textwidth]{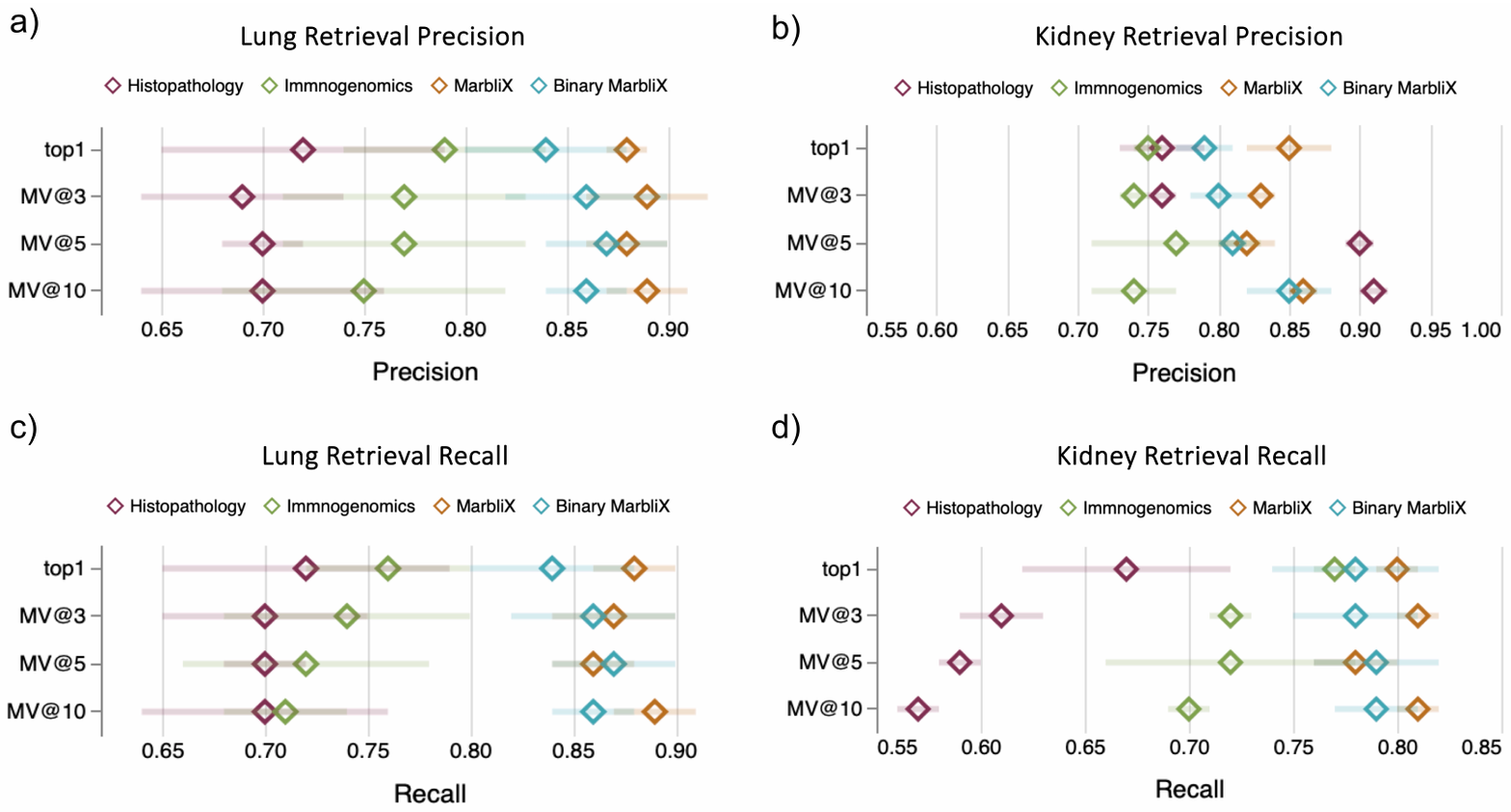}
        \caption{Multimodal search performance of real and binary MarbliX representations vs. unimodal image and immunogenomic embeddings for (a, c) lung and (b, d) kidney datasets. Diamonds indicate mean macro-average precision and recall; error bars show standard deviation. Retrieval is based on top-1 and MV@3/5/10 using leave-one-out and majority vote.}
        \label{fig:results2}
    \vspace{0.5cm}
\end{figure}

MarbliX’s monogram representation enables efficient multimodal case search and retrieval (Figure \ref{fig:multimodal_search}). Each patient's histopathology and immunogenomic data are fused into a single representation. A leave-one-out validation on the test folds (not used in training) was performed. Each test case served as a query against the monogram archive, using a majority vote on the top-3, top-5, and top-10 (MV@3, MV@5, MV@10) retrieved cases to assess performance. Figure \ref{fig:results2} shows the macro average F1-score and accuracy for lung and kidney datasets. In lung (Figures \ref{fig:results2}(a) and (c)), MarbliX achieves 85–89\% across all measures, outperforming histopathology (69–71\%) and immunogenomics (73–76\%). For kidney (Figures \ref{fig:results2}(b) and (d)), real monograms perform best (F1: 80–83\%, Accuracy: 87–90\%), with binary monograms slightly lower (F1: 78–82\%). Immunogenomics outperforms histopathology in F1 (70–76\% vs. 60–70\%), with comparable accuracy. All kidney representations show lower F1 than accuracy due to class imbalance. Precision and recall results (Figure \ref{fig:results2}) further highlight MarbliX's consistency, with shorter standard deviation bars indicating stable performance across folds. MarbliX outperforms unimodal representations in both precision and recall. For kidney, histopathology yields the highest precision for MV@5 and MV@10 (90–91\%) but the lowest recall (56–59\%). Binary MarbliX matches immunogenomics in recall for top-1, but with higher precision. Overall, MarbliX maintains both precision and recall above 78\% across all evaluation criteria.

\subsection*{MarbliX Monogram Binary Representation Analysis}
\begin{figure}
    \centering   
    \vspace{-10pt}
    \setlength{\abovecaptionskip}{2pt} 
    \setlength{\belowcaptionskip}{-6pt} 
\includegraphics[width=\textwidth]{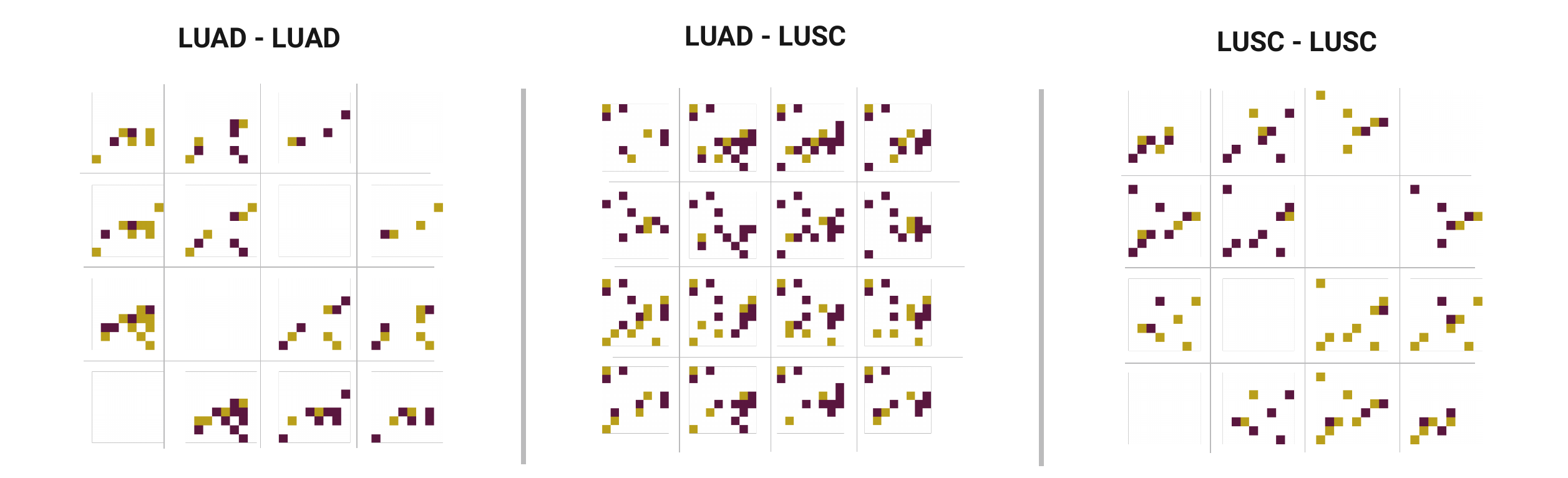}
        \caption{Monogram representations generated using MarbliX, for randomly selected patients with LUAD LUSC. The figure presents matrix bitwise XOR results.
        In the context of $\{\text{LUAD}_{set} \oplus \text{LUSC}_{set}\}$, yellow pixels signify features unique to LUAD matrices, absent in LUSC, and purple pixels denote features specific to LUSC matrices, absent in LUAD.}
        \label{fig:lung_matrices}
\end{figure}

MarbliX aims to generate multimodal \emph{monogram} representations that are similar for patients with the same cancer subtype and distinct for those with different subtypes. Figure \ref{fig:lung_matrices} illustrates this, showing four LUAD and four LUSC monograms for randomly selected samples (quantified in Figure \ref{fig:results1}(d)). Monograms from patients within the same subtype display consistent patterns (intra-similarity). To highlight this, each LUAD matrix was XORed with the others in its set, and the same was done for LUSC (${\text{LUAD}{set} - \text{LUAD}{set}}$, ${\text{LUSC}{set} - \text{LUSC}{set}}$). Cross-subtype dissimilarity was assessed by XORing LUAD with LUSC (${\text{LUAD}{set} - \text{LUSC}{set}}$). In the resulting matrices, yellow pixels indicate a 0-to-1 change, and purple a 1-to-0 change. This color coding clarifies which features are unique to each subtype. As shown in Figure \ref{fig:lung_matrices}, intra-subtype differences are smaller (with more white space), while inter-subtype differences are greater (denser yellow and purple areas), demonstrating MarbliX’s ability to distinguish between LUAD and LUSC representations.

\subsection*{\textcolor{black}{Settings and Ablation}}
\textcolor{black}{Triplet construction, normalization, and random seeds follow standard protocols. The preprocessing filters (e.g., removing extremely rare sequences) were applied globally across the dataset, not per class, and therefore cannot introduce label leakage. Architectural settings—such as the 128-dimensional bottleneck, 8×8 monogram size, and autoencoder layer widths—were chosen as pragmatic defaults balancing stability and compactness, and preliminary checks indicated that moderate variations did not change overall retrieval trends. A full sensitivity analysis is outside the scope of this initial study. We also did not provide full ablation studies on architectural parameters (e.g., bottleneck size, monogram dimensionality, autoencoder depth) or preprocessing thresholds. These choices were selected as practical defaults rather than the result of exhaustive tuning, and while preliminary tests suggested robustness to moderate variation, a systematic analysis was beyond our scope. We acknowledge that preprocessing thresholds were not explored for sensitivity. Importantly, all filtering operations were applied globally, avoiding any risk of class-wise information leakage. As an \textbf{ablation} experiment, we run multiple experiments  to  compare unimodal WSI features, unimodal immunogenomic features, non-binarized MarbliX representations, and binarized MarbliX monograms. Figure shows the results. }

\begin{figure}
    \vspace{-23pt}
    \centering
\includegraphics[width=1\textwidth]{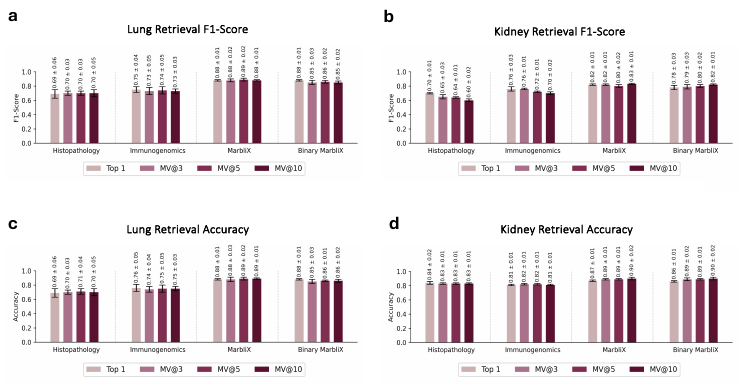}
    \caption{\textcolor{black}{While unimodal features exhibit moderate subtype separability, integrating modalities through our shared latent space notably improves classification performance. Furthermore, binarizing the shared latent vectors into 8×8 monograms preserves class structure with minimal loss in discriminability, demonstrating their utility as compact yet informative representations.}}
    \label{fig:ablation}
\end{figure}
\section{Discussion} 
This study addresses the limited exploration of integrating histopathology images with immunogenomic data—a combination with significant potential in cancer research. The proposed MarbliX framework aims to bridge this gap by enabling unified, multimodal patient representations that support deeper insights and novel research directions. 


\textcolor{black}{For biological interpretability, MarbliX can present case-level clinical and histologic comparisons by displaying representative “neighbor” cases retrieved via the multimodal monogram. For example, a HER2-enriched breast cancer case with high lymphocytic infiltration \cite{schettini2020her2} would retrieve neighbors showing similar immune-dense stroma and comparable HER2 expression profiles. Likewise, a low-grade colorectal adenocarcinoma with MSI-high status \cite{schrock2019tumor} would be paired with cases exhibiting matching glandular morphology and parallel mismatch-repair signatures. These intuitive cross-modal pairings can help clinicians verify that retrieved samples align with both the clinical phenotype and microscopic appearance.}

Experimental results show that MarbliX effectively captures distinguishing features from both histopathology and immunogenomic data. Similarity analyses revealed consistent intra-class patterns and distinct inter-class differences. t-SNE visualizations further demonstrated its strength in subtype separation. By converting complex multimodal data into binary matrices, MarbliX supports efficient, interpretable integration of patient information, aiding data-driven decision-making in both clinical and research contexts.

\textbf{Broader Impacts:} The proposed MarbliX framework offers significant societal benefits by enhancing personalized diagnostics and accelerating research through efficient multimodal data integration. By compressing patient data into binary formats, it supports scalable and interpretable comparisons for clinical decision-making and research.
However, risks remain. Unrepresentative training data may introduce bias, and overreliance on automated suggestions could reduce necessary human oversight. To address this, rigorous data curation and maintaining human-in-the-loop oversight are essential for clinical use.

\textbf{Limitations:} MarbliX shows strong performance in generating multimodal patient representations, but several considerations remain. Preprocessing must be tailored to each modality (e.g., histopathology, immunogenomics), and its effectiveness depends on how well data are embedded into a shared space. Performance can also be affected by input quality, such as low-resolution slides or incomplete genomic profiles. While scalable and efficient, using large language models for feature extraction can be computationally intensive. 

\textcolor{black}{\textbf{Deep embeddings} inevitably collide because compressing complex tissue morphology into a fixed-size vector forces many distinct cases into overlapping regions. They are also not fully stable—small perturbations, rotations, stains, or artefacts can shift embeddings unpredictably—and their finite capacity cannot capture the combinatorial diversity of pathology. As datasets grow, embeddings become crowded, degrading fine-grained diagnosis and retrieval. Bimodal patient codes avoid this by keeping modality-specific channels, reducing information loss and collisions while preserving structure and interpretability. They scale better for retrieval and remain more stable, since perturbations in one modality do not distort the entire representation.}

\subsection{Limitations}
\textcolor{black}{This work does not include direct comparisons with standard multimodal fusion baselines (e.g., CLIP-style contrastive models, BEiT/BLIP, co-attention transformers, or concatenation-based classifiers). These methods rely on large paired datasets and continuous text or transcriptomic embeddings, which are not available or directly compatible with sparse, sequence-derived immunogenomic features. Adapting such architectures to WSI–immune-repertoire integration would require substantial redesign of both encoders and fusion modules. Our focus was therefore on evaluating the monogram as an efficient indexing representation rather than benchmarking alternative multimodal fusion strategies.}

\section*{Data and Code Availability}
The code is available at \url{https://github.com/KimiaLabMayo/MarbliX}. The datasets used in this study are obtained from TCGA and can be obtained through their respective portals.


\newpage
\bibliographystyle{splncs04}
\bibliography{sn-bibliography}
\end{document}